# Strong antiferromagnetic interaction owing to a large trigonal distortion in the spin-orbit-coupled honeycomb lattice iridate CdIrO$_3$

Yuya Haraguchi[†], and Hiroko Aruga Katori
Department of Applied Physics and Chemical Engineering, Tokyo University of Agriculture and Technology, Koganei, Tokyo 184-8588, Japan
[†]chiyuya3@go.tuat.ac.jp

We investigated the magnetic properties of the ilmenite-type iridate CdIrO$_3$ with a honeycomb lattice formed by Ir$^{4+}$ ions prepared via a solid-state metathesis. The magnetization measurements with using the powder sample reveal a large effective magnetic moment and a fairly strong antiferromagnetic interaction, indicating a deviation from the Kitaev model. Considering the relationship between magnetism and crystal structure in CdIrO$_3$ with comparing with the other ilmenite-type iridates ZnIrO$_3$ and MgIrO$_3$, we conclude that insulating CdIrO$_3$ cannot be describe as a $J_{\text{eff}} = 1/2$ Mott state owing to a metathetically-stabilized large trigonal distortion of IrO$_6$ octahedra.

## I. Introduction

Recently, physical properties driven by the spin-orbit coupling (SOC) have attracted much attention from theorists and experimentalists [1-10]. An electronic state of a magnetic ion under the strong spin-orbit coupling is well described by $J_{\text{eff}}$ pseudospins formed by the combination of SOC and orbital degeneracy. In the limit of strong electron correlation, unconventional magnetic interactions are theoretically predicted to generate among $J_{\text{eff}}$ pseudospins, which results in the realization of unconventional quantum ground states [10-12]. In particular, it is known that a Kitaev-type bond-directional highly anisotropic ferromagnetic interaction is realized between spin-orbital coupled $J_{\text{eff}} = 1/2$ electrons, which is explained by the Jackeli-Khaliullin mechanism [10]. In the situation of presence of the Kitaev-type interaction on the honeycomb lattice, the ground state is exactly solved to be a quantum spin liquid [2]. Under these circumstances, searching for honeycomb lattice magnets formed by 4$d$/5$d$ transition metal ions with a $d^5$ electron configuration is becoming an active area of research. In fact, some signs of Kitaev spin liquid have been found in some realistic compounds—α-RuCl$_3$ and H$_3$LiIr$_2$O$_6$ [13-16]. In the case of α-RuCl$_3$, a strong evidence of Kitaev spin-liquid is being found as a half-integer thermal quantum Hall effect [15]. H$_3$LiIr$_2$O$_6$ shows a spin-liquid behavior as a ground state [16]. The consequences of the Kitaev interaction, however, are not completely understood.

In the Kitaev-type interaction as described above, the $J_{\text{eff}} = 1/2$ state assumes a local cubic symmetric field with a perfect $M$O$_6$ octahedron. All realistic compounds, however, have nonideal octahedra deviated from a cubic crystal field. It has been found in some iridium oxides that the $J_{\text{eff}} = 3/2$ state is mixed with the ground state $J_{\text{eff}} = 1/2$ state due to the trigonal/tetragonal distortion [17-19]. Because of an admixture of $J_{\text{eff}} = 3/2$ component, the ground state cannot be described by the pure $J_{\text{eff}} = 1/2$ wave function. It is not known in detail how such admixture of wave functions affects the Kitaev-type interactions. Therefore, it is important to compare the relationship between magnetism and local crystal distortion in various $d^5$ honeycomb lattice magnets because it is naturally not a cubic symmetry field.

In this paper, we report on the successfully synthesis of a new ilmenite-type honeycomb lattice iridate CdIrO$_3$ via the metathesis reaction as well as its magnetic properties. Since the crystal structure of CdIrO$_3$ is qualitatively the same as MgIrO$_3$ and ZnIrO$_3$ [20], we can use the new material to systematically study the effect of lattice distortion of local crystal field on the Kitaev magnetism. Indeed, the observed magnetic behavior could not be explained in the $J_{\text{eff}} = 1/2$ state, which would be cause by a large trigonal distortion probed by the analysis of crystal structure. CdIrO$_3$ is expected to be a good model compound for clarifying the effect of a local distortion in the physics of spin-orbital-entangled Mott insulators

## II. Experimental Methods

We designed the synthesis route of CdIrO$_3$ by modifying the metathesis synthesis method of MgIrO$_3$ and ZnIrO$_3$ [16] as follows:

$$\text{Li}_2\text{IrO}_3 + \text{CdCl}_2 \rightarrow \text{CdIrO}_3 + 2\text{LiCl}. \quad (1)$$

The precursor Li$_2$IrO$_3$ was obtained by the conventional solid-state reaction method according to the previous report [21]. This precursor was ground well with 5-fold excess of $CdCl_2$ in an Ar-filled glovebox, sealed in an evacuated Pyrex tube, and reacted at 350 or 400°C. In both conditions, however, $CdIrO_3$ could not be synthesized: a mixture at 350°C is unreactive, and a mixture at 400°C completely decomposes into $IrO_2$ and some unspecified impurities. In order to lower the reaction temperature and to suppress decomposition, an inert salt of NaCl is added to a reaction mixture before calcination at 350°C for 100 h. We found that this process is important for a stabilization of metastable $CdIrO_3$. The unreacted starting material $CdCl_2$, the inert salt NaCl, and the byproduct LiCl were removed by washing the sample with distilled water. The product was characterized by powder X-ray diffraction (XRD) experiments in a diffractometer with Cu-Kα radiation. The cell parameters and crystal structure were refined by the Rietveld method using the RIETAN-FP v2.16 software [22].

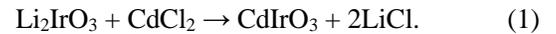

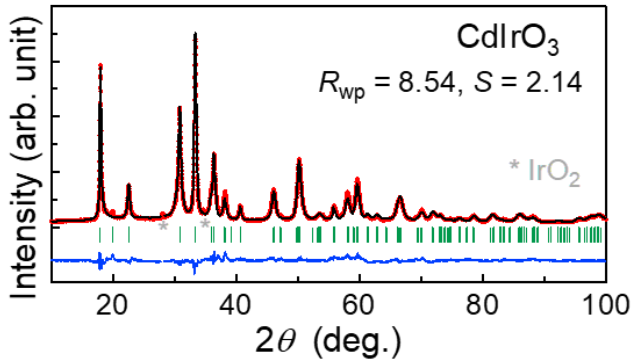

**Fig. 1** Powder x-ray diffraction patterns of $CdIrO_3$. The observed intensities (red), calculated intensities (black), and their differences (blue) are shown. Vertical bars indicate the positions of Bragg reflections. The peak marked with an asterisk is an impurity peak due to unwashed $IrO_2$ remaining in the final product.

**TABLE I** Crystallographic parameters for $CdIrO_3$ ($R\bar{3}$) determined from powder x-ray diffraction experiments. The obtained lattice parameters are $a = 5.3679(3)$, $c = 14.8108(3)$ Å. $B$ is the atomic displacement parameter.

|    | site | $x$       | $y$         | $z$       | $B$ (Å$^2$) |
|----|------|-----------|-------------|-----------|-------------|
| Cd | 6$c$ | 0         | 0           | 0.3675(1) | 0.15        |
| Ir | 6$c$ | 0         | 0           | 0.1614(1) | 0.87        |
| O  | 18$f$| 0.3473(23)| -0.0442(21) | 0.1205(6) | 1.3         |

The temperature dependence of the magnetization of powder samples was measured under several magnetic fields up to 7 T by using a magnetic property measurement system (MPMS; Quantum Design) equipped at the Institute for Solid State Physics at the University of Tokyo.

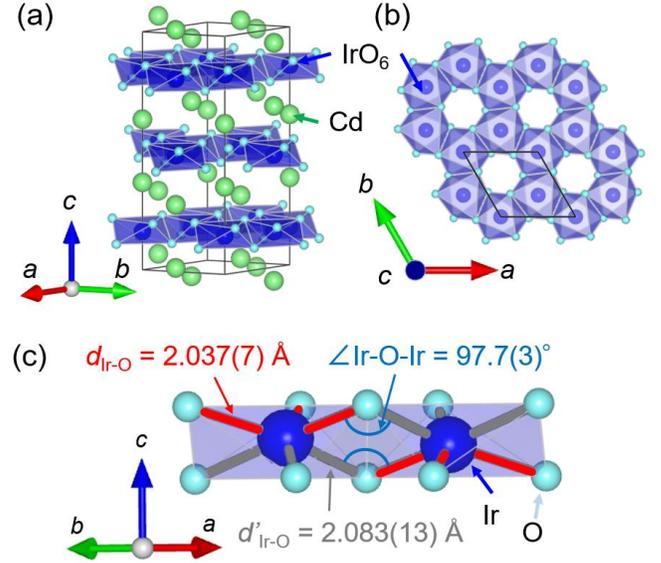

**Fig. 2** Crystal structure of $CdIrO_3$ (a) viewed along the $c$-axis and (b) perpendicular to the $ab$-plane. (c) The local environment of a pair of edge-sharing $IrO_6$ octahedra with the two different bond lengths of Ir-O and the bond angle of Ir-O-Ir. The VESTA program is used for visualization [22].

**TABLE II** Ir-O bond lengths (Å) and Ir-O-Ir bond angle (°) in $CdIrO_3$ obtained from the powder x-ray diffraction data (including those of $MgIrO_3$ and $ZnIrO_3$ for comparison [20]).

| Bond          | $CdIrO_3$  | $MgIrO_3$ | $ZnIrO_3$ |
|---------------|------------|-----------|-----------|
| Ir-O (×3)     | 2.037(7)   | 1.942(6)  | 1.990(3)  |
| Ir-O (×3)     | 2.083(13)  | 2.136(9)  | 2.068(6)  |
| Bond angle    | $CdIrO_3$  | $MgIrO_3$ | $ZnIrO_3$ |
| Ir-O-Ir       | 97.7(3)    | 94.0(3)   | 95.7(1)   |

### III. Results

Powder x-ray diffraction pattern of $CdIrO_3$ is shown in Fig. 1. All the peaks except for those from impurity of a trace amount of $IrO_2$ can be indexed by the ilmenite-type structure with the space group of $R\bar{3}$ with the lattice constants $a = 5.3679(3)$ Å and $c = 14.8108(3)$ Å, which are more expanded than that of $ZnIrO_3$ and $MgIrO_3$. As shown in Fig. 2(a) and 2(b), Ir ions form a regular honeycomb lattice in the ilmenite structure. The structure of $CdIrO_3$ is refined with using the Rietveld method as described in the experimental section. The detail of the refinement parameters is given in Table 1. The bond valence sum calculation for Ir ions from the refined structural parameters [see TABLE II and Fig. 2(c)] yields +3.994, which is consistent with the expected valence of +4.

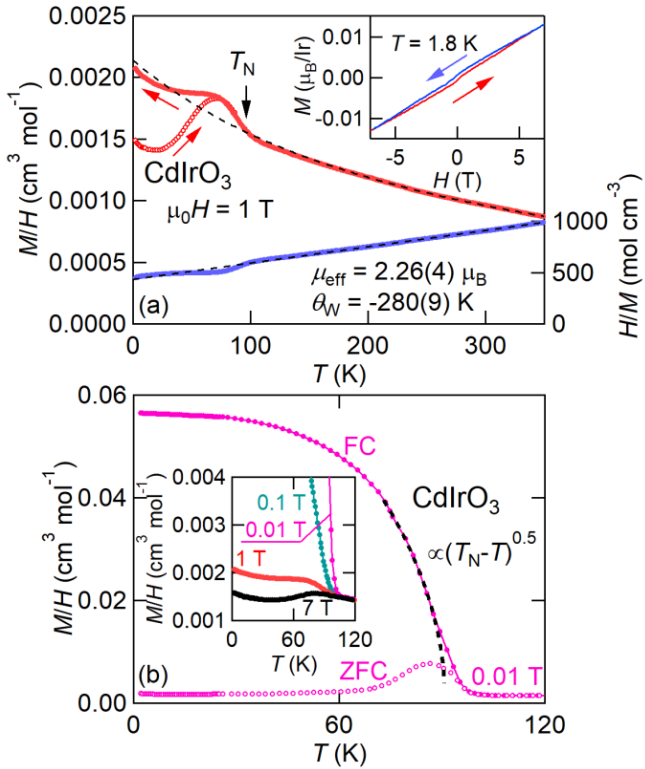

**Fig. 3** (a) Temperature dependences of magnetic susceptibility $M/H$ and its inverse of powder samples $CdIrO_3$ under a magnetic field of 1 T. The measurements were conducted upon heating after zero field-cooling (ZFC) and then upon cooling (FC), as shown by the arrows. The dashed line is a result of Curie-Weiss fitting. The values are those of estimated magnetic interactions. The inset shows the isothermal magnetization curve measured at $T = 1.8$ K for $CdIrO_3$. (b) ZFC and FC $M/H$ curves of $CdIrO_3$ under low magnetic field $\mu_0 H = 0.01$ T. The black dashed line is the result of fitting near $T_N$ as described in the text. The inset shows temperature dependences of FC $M/H$ curves measured for several magnetic fields.

An ilmenite type structure is another structural type of composition $AMO_3$ along with perovskite. With using the ratio of ion radius $r(A^{2+})/r(O^{2+})$, one can simply estimate which structure type is stable: an ilmenite is stable in $r(A^{2+})/r(O^{2+}) < 0.7$, while a perovskite in $r(A^{2+})/r(O^{2+}) > 0.7$ [23]. In ilmenite $CdIrO_3$, the value of $r(Cd^{2+})/r(O^{2+}) = 0.688$ satisfies the criterion, which is in contrast to the ratio $r(Ca^{2+})/r(O^{2+}) = 0.72$ in post-perovskite $CaIrO_3$ [24]. The bond angles of Ir-O-Ir in $CdIrO_3$ is found as 97.7(3)°, which is larger than those in $ZnIrO_3$ (95.7(1)°) and $MgIrO_3$ (94.0(3)°) [see TABLE II and Fig. 2(c)]. The difference in the angle of the super-exchange path is thought to affect the spin model as described later.

The temperature dependence of magnetic susceptibility $M/H$ and its inverse $H/M$ for $CdIrO_3$ measured under 1 T are shown in Fig. 3. There is a linear relationship in $H/M$ versus $T$ in the high temperature region, indicating the presence of local magnetic moment. A Curie-Weiss fitting of $H/M$ at 250-350 K yields an effective magnetic moment $\mu_{eff} = 2.26(4)$ $\mu_B$ and Weiss temperature $\theta_W = -280(9)$ K. The effective moment is larger than the expected value of 1.73 for $J_{eff} = 1/2$. The origin of large deviation is discussed later. The large negative value of $\theta_W$ indicates predominantly antiferromagnetic interaction between the $Ir^{4+}$ ions.

At low temperature below approximately 100 K, $M/H$ curve starts to increase as well as a thermal hysteresis between the zero-field-cooled (ZFC) and field-cooled (FC) data, indicating magnetic order. In Fig. 3(b), the $M/H$ curve of $CdIrO_3$ measured under 0.01T is plotted as a function of $T$. Below approximately 100 K, FC $M/H$ starts to increase rapidly and shows saturation behavior as the temperature is decreased. In addition, a thermal hysteresis between ZFC and FC under 0.01 T becomes more apparent than that under 1 T As shown in the inset of Fig. 3(b), with increasing of magnetic field, the increase of FC $M/H$ is suppressed. These behaviors indicate the presence of ferromagnetic moment. In the category of molecular field approximation, a spontaneous magnetization $M$ shows a critical behavior of $M \propto (T-T_N)^{0.5}$ below $T_N$. With using the function, the critical temperature is roughly estimated as $T_N = 90.9(1)$ K. Just above $T_N$, magnetization is larger than the curve of critical behavior, which is due to the effect of magnetic field. Under higher magnetic fields 7 T, $M/H$ has a kink near $T_N$, which is similar to the $M/H$ of a material with conventional antiferromagnetic ordering. Thus, it is reasonable to think that a canted antiferromagnetic order with a weak ferromagnetic moment occurs at $T_N$. As shown in the inset of Fig. 3(a), a magnetic hysteresis loop is observed in the isothermal magnetization at 1.8 K, which also demonstrates the presence of a weak ferromagnetic moment. Such a weak ferromagnetism has been observed also in $MgIrO_3$ [20].

## IV. Discussion

The observed large $\mu_{eff}$ in CdIrO$_3$ would suggest that the electronic state is deviated from an ideal $J_{eff} = 1/2$ state. Here, we discuss the relationship between the local electronic state and the crystal field. As shown in Fig. 4(a), in a cubic crystal field, three degenerated $t_{2g}$ orbitals split into the $J_{eff} = 3/2$ quartet and $J_{eff} = 1/2$ doublet by SOC. The local electronic state, however, is sensitive to the trigonal distortion of the IrO$_6$ octahedra. Without SOC, the threefold degenerate $t_{2g}$ level splits into twofold degenerate higher levels $e_g$ and a nondegenerate lower level $a_{1g}$ by a trigonal distortion. Thus, when both a trigonal distortion and SOC are included, the $t_{2g}$ levels split into three Kramers doublets with the degeneracy fully lifted as shown in Fig 4(b). When a trigonal crystal field is large comparable to SOC, the ground state $\varphi_0$ derived from $J_{eff} = 1/2$ should be apart from the pure $J_{eff} = 1/2$ wave function because of an admixture of $J_{eff} = 1/2$ and $J_{eff} = 3/2$ component. Thus, the magnetism is sensitive to the degree of distortion of the octahedron. Indeed, previous investigations of a resonant inelastic x-ray scattering (RIXS) detect the admixture of $J_{eff} = 1/2$ and $J_{eff} = 3/2$ states in some iridates [17-19]. In order to reveal the detail of ground electronic state in $A$IrO$_3$, it is necessary to conduct the RIXS measurement and it is a future issue.

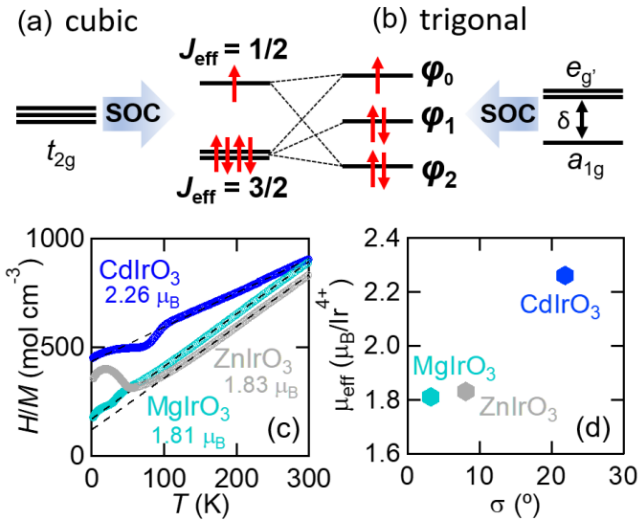

**Fig. 4** A schematic view of (a) splitting of $J_{eff} = 1/2$ and $3/2$ states under cubic crystal field into (b) three Kramers doublets $\varphi_0$, $\varphi_1$ and $\varphi_2$ under trigonal crystal field. (c) The temperature dependence of inversed magnetic susceptibility $H/M$ in three ilmenite-type iridates. (d) Relationship between effective magnetic moments $\mu_{eff}$ and bond angle variances $\sigma$ in three ilmenite-type iridates.

The degree of trigonal distortion in the IrO$_6$ octahedra can be parameterized by the bond angle variance [25],

$$\sigma = \sqrt{\sum_{i=1}^{12} \frac{(\varphi_i - \varphi_0)^2}{m-1}}, \quad (2)$$

where $m$ is the number of anion-cation-anion bond angles, $\phi_i$ is the $i$-th bond angle of the distorted coordination-polyhedra, and $\phi_0$ is the bond angle of the coordination-polyhedra with $O_h$ symmetry. In the case of octahedra, $\phi_0$ is 90°. The value of $\sigma$ in CdIrO$_3$ is found as 21.9°, which is significantly larger than that of ZnIrO$_3$ (8.05°) and MgIrO$_3$ (3.20°). Thus, the IrO$_6$ octahedra in CdIrO$_3$ yields significantly larger trigonal distortion than that in ZnIrO$_3$ and MgIrO$_3$. Let us recall that the observed $\mu_{eff}$ of 2.26 $\mu_B$/Ir in CdIrO$_3$ is significantly larger than 1.73 $\mu_B$ in pure $J_{eff} = 1/2$. The deviation of $\mu_{eff}$ from the ideal value of 1.73 should be a parameter of mixing $J_{eff} = 1/2$ and $J_{eff} = 3/2$. As shown in Fig. 4(c) and 4(d), one finds that the value of $\mu_{eff}$ tends to increase with increasing $\sigma$, indicating a presence of strong correlation between a degree of trigonal distortion and a magnitude of mixing $J_{eff} = 1/2$ and $J_{eff} = 3/2$. From these facts indicate that the electronic state of CdIrO$_3$ cannot be describe as a pure $J_{eff} = 1/2$ iridate.

The absolute value of $\theta_W = -280$ K in CdIrO$_3$ is extremely larger than those in MgIrO$_3$ (−48 K) and ZnIrO$_3$ (−66 K) [20]. Therefore, this large negative Weiss temperature means that the ferromagnetic Kitaev term is much smaller than the antiferromagnetic Heisenberg term. The observation of high temperature magnetic transition $T_N = 90$ K also evidences that the Kitaev term is much smaller than the primary Heisenberg term in CdIrO$_3$. It is theoretically known that a deviation of Ir-O-Ir angle from 90° enhances the antiferromagnetic Heisenberg term [10]. Indeed, the Ir-O-Ir angle of CdIrO$_3$ is larger compared to the angles of MgIrO$_3$ and ZnIrO$_3$, which consistent with higher $\theta_W$ and $T_N$ in CdIrO$_3$. In a mixed state of $J_{eff} = 1/2$ and $3/2$, the cancellation of antiferromagnetic Heisenberg term would be lifted, resulting in that ferromagnetic Kitaev term should become relatively smaller. Thus, a mixing of $J_{eff} = 1/2$ and $3/2$ states, which is indirectly probed by the larger $\mu_{eff}$, has a potential to be another origin of relatively small Kitaev term.

We conclude that CdIrO$_3$ is *not* a pure $J_{eff} = 1/2$ iridate. Therefore, the observed weak ferromagnetism would not be due to the Kitaev interaction. Such a weak ferromagnetism has been observed also in a similar ilmenite-type manganate ZnMnO$_3$ without SOC [26]. Thus, it is reasonable to think that the dominant origin of these weak ferromagnetism is a Dzyaloshinskii-Moriya (DM) interaction. In an ilmenite structure, there is a ***D***-vector not on the nearest-neighbor interaction ($J_1$) but on the next-nearest-neighbor one ($J_2$) in the honeycomb layer because of the presence/absence of inversion symmetry. DM interaction is proportional to the magnitude of superexchange interaction. Thus, when DM interaction on $J_2$ is effective, it is considered that $J_2$ is a sufficiently large. Indeed, in $CdIrO_3$ the frustration index $f = |\theta_W|/T_N$ is 3.08, suggesting that the magnetic ordering is suppressed by the spin frustration in the spin model of $J_1$-$J_2$ honeycomb lattice.

It is experimentally found that less distortion of $IrO_6$ octahedra is a key to realize a pure Kitaev model. As a guideline for suppressing distortion, we propose two strategies. One strategy is to synthesis a hypothetical ilmenite iridium oxide $A$IrO$_3$ with a smaller $A$ ion, for example, $BeIrO_3$. However, since $Be^{2+}$ is known to prefer tetrahedral coordination rather than octahedral one, an ilmenite type $BeIrO_3$ may not be able to synthesize. The other strategy is an application of high pressure to tune the lattice constant. In fact, the distortion parameter σ tends to decrease with the lattice constant shrinking. Therefore, in a high-pressure experiment of $MgIrO_3$ with the least trigonal distortion in ilmenite iridates, there is a possibility to approach a pure Kitaev model to exhibit a quantum spin liquid.

## V. Summary

We have successfully synthesized a metastable honeycomb lattice iridate $CdIrO_3$ with an ilmenite structure via a metathesis reaction and investigated its crystal structure and magnetism. The observed effective magnetic moment 2.26 $\mu_B$/Ir is larger than that in pure $J_{eff}$ = 1/2 states. Considering the relationship between the crystal structure and the magnetism, it is reasonable to think that the ground state of $CdIrO_3$ cannot be described as the $J_{eff}$ = 1/2 state owing to a large trigonal distortion. The large negative Weiss temperature also supports the deviation from the pure Kitaev model. These results experimentally show that the distortion of local crystal field counteracts the realization of pure Kitaev model in realistic materials. Therefore, the effect of the local crystal distortion should not be dismissed in the local physics of spin-orbital-entangled Mott insulators.

## Acknowledgement

This work was supported by Japan Society for the Promotion of Science (JSPS) KAKENHI Grant Number JP19K14646 and JP18K03506. Part of this work was carried out by the joint research in the Institute for Solid State Physics, the University of Tokyo.

## Reference

[1] W. Witczak-Krempa, G. Chen, Y.B. Kim, and L.Balents, Annu. Rev. Condens. Matter Phys. **5**, 57 (2014).
[2] A. Kitaev, Ann. Phys. **321**, 2 (2006).
[3] X.Wan, A.M.Turner, A.Vishwanath, and S.Y. Savrasov,Phys. Rev. B **83**, 205101 (2011).
[4] Y. Okamoto, M. Nohara, H. A. Katori, and H. Takagi, Phys. Rev. Lett. **99**, 137207 (2007).
[5] M. J. Lawler, A. Paramekanti, Y. B. Kim, and L. Balents, Phys. Rev. Lett. **101**, 197202 (2008).
[6] A. Shitade, H. Katsura, J. Kunes, X.-L. Qi, S.-C. Zhang, and N. Nagaosa, Phys. Rev. Lett. **102**, 256403 (2009).
[7] B. J. Kim, H. Ohsumi, T. Komesu, S. Sakai, T. Morita, H. Takagi, and T. Arima, Science **323**, 1329 (2009).
[8] B. J. Kim, Hosub Jin, S. J. Moon, J.-Y. Kim, B.-G. Park, C. S. Leem, Jaejun Yu, T. W. Noh, C. Kim, S. -J. Oh, J. -H. Park, V. Durairaj, G. Cao, and E. Rotenberg, Phys. Rev. Lett. **101**, 076402 (2008).
[9] W. Witczak-Krempa and Y.-B. Kim, Phys. Rev. B **85**, 045124 (2012).
[10] G. Jackeli and G. Khaliullin, Phys. Rev. Lett. **102**, 017205 (2009).
[11] G. Khaliullin, Phys. Rev. Lett. **111**, 197201 (2013).
[12] M. G. Yamada, M. Oshikawa, and G. Jackeli, Phys. Rev. Lett. **121**, 097201 (2018).
[13] A. Banerjee, C. A. Bridges, J.-Q. Yan, A. A. Aczel, L. Li, M. B. Stone, G. E. Granroth, M. D. Lumsden, Y. Yiu, J. Knolle, S. Bhattacharjee, D. L. Kovrizhin, R. Moessner, D. A. Tennant, D. G. Mandrus, and S. E. Nagler, Nat. Mater. **15**, 733 (2016).
[14] S.-H. Baek, S.-H. Do, K.-Y. Choi, Y. S. Kwon, A. U. B. Wolter, S. Nishimoto, J. van den Brink, and B. Buchner, Phys. Rev. Lett. **119**, 037201 (2017).
[15] Y. Kasahara, T. Ohnishi, Y. Mizukami, O. Tanaka, S. Ma, K. Sugii, N. Kurita, H. Tanaka, J. Nasu, Y. Motome, T. Shibauchi, and Y. Matsuda, Nature (London) **559**, 227 (2018).
[16] K. Kitagawa, T. Takayama, Y. Matsumoto, A. Kato, R. Takano, Y. Kishimoto, S. Bette, R. Dinnebier, G. Jackeli, and H. Takagi, Nature (London) **554**, 341 (2018).
[17] M. Moretti Sala, K. Ohgushi, A. Al-Zein, Y. Hirata, G. Monaco, and M. Krisch, Phys. Rev. Lett. **112**, 176402 (2014).
[18] X. Liu, Vamshi M. Katukuri, L. Hozoi, Wei-Guo Yin, 1 M. P. M. Dean, M. H. Upton, Jungho Kim, D. Casa, A. Said, T. Gog, T. F. Qi, G. Cao, A. M. Tsvelik, Jeroen van den Brink, and J. P. Hill, Phys. Rev. Lett. **109**, 157401 (2012).
[19] J. Sheng, F. Ye, C. Hoffmann, V. R. Cooper, S. Okamoto, J. Terzic, H. Zheng, H. Zhao, and G. Cao, Phys. Rev. B **97**, 235116 (2018).
[20] Y. Haraguchi, C. Michioka, A. Matsuo, K. Kindo, H.

Ueda, and K. Yoshimura, Phys. Rev. Mater. **2**, 054411 (2018).
[21] I. Felner and I.M. Bradaric, Physica (Amsterdam) **311B**, 195 (2002).
[22] F. Izumi and K. Momma, Solid State Phenom. **130**, 15 (2007).
[23] Ulrich Muller, *Inorganic Structural Chemistry*, *2$^{nd}$ Edition* (John Wiley and Sons Ltd, 2007).
[24] K. Ohgushi, J. -I. Yamaura, H. Ohsumi, K. Sugimoto, S. Takeshita, A. Tokuda, H. Takagi, M. Takata, and T. -H. Arima, Phys. Rev. Lett. **110**, 217212 (2013).
[25] K. Robinson, G. V. Gibbs, P. H. Ribbe, Science **172**, 567 (1971).
[26] Y. Haraguchi, K. Nawa, C. Michioka, H. Ueda, A. Matsuo, K. Kindo, M. Avdeev, T. J. Sato, and K. Yoshimura, Phys. Rev. Mater. **3**, 124406 (2019).